
\def\begin#1{\vskip .25in\goodbreak\noindent
		    {\bf{#1}}}

\def\beginsection#1{\vskip .25in\goodbreak\noindent
		    {\bf{#1}}\nobreak
		    \smallskip\nobreak\noindent\hskip -.0mm}


\def\al{\alpha}
\def\be{\beta}
\def\ga{\gamma}

\def\th{\theta}

\def\om{\omega}

\def\W{$W$-algebra}
\def\wthree{$W_3$-algebra}
\def\wfour{$W_4$-algebra}
\def\vir{Virasoro}
\def\km{Kac-Moody}
\def\mc{Maurer-Cartan equations}
\def\fda{free differential algebra}
\def\half{{\textstyle{1\over2}}}
\def\gfc{Gel'fand-Fuchs cocycle}
\def\intx{\int dx\,}
\def\n#1{\nu^{(#1)}}
\def\m#1{\mu^{(#1)}}
\def\gdo{Gel'fand-Dickey operator}
\def\cs{Chern-Simons}
\def\psip{\psi_+}
\def\psim{\psi_-}
\def\D{\bar D}
\def\fr#1/#2.{{\textstyle{#1\over#2}}}

\def\dJ{\dot J}
\def\intxth{\intx d\bar\theta\,d\theta\,\int}
\def\letter{letter}
\def\3{three dimension}
\def\eom{equations of motion}

\pageno=0
\vskip 1in
\centerline{\bf DUAL FORMULATION OF CLASSICAL W-ALGEBRAS}
\centerline{\bf }

\vskip 1.in
\centerline{R.E.C. Perret}
\smallskip

\vskip .5cm
\centerline{\it Department of Physics}
\centerline{\it University of Southern California}
\centerline{\it Los Angeles, CA 90089--0484}
\vskip 1.2in
\centerline{Submitted to: \it Letters in Mathematical Physics}

\vskip 1.2in

\noindent{\bf Abstract:}\hfill\break\noindent
By extending the concept of \mc, I introduce a dual formulation of
(classical) nonlinear extensions of the \vir\ algebra. This dual
formulation is closely related to three dimensional actions
which are analogous to a \cs\ action.
I present an explicit construction in terms of superfields of the $N=2$
super \wfour.
\vskip1.5cm\noindent USC--92/16\hfill August 1992
\vskip 1in
\eject

\def\ZFL{1}
\def\GD{2}
\def\cw{3}
\def\CB{4}
\def\CDF{5}
\def\ik{6}
\def\HV{7}
\def\SvN{8}
\def\P{9}
\def\LPRSW{10}
\def\SVN{11}

\begin{1.} \W s [\ZFL] are nonlinear extensions of the \vir\ algebra
which play an important role in many areas in mathematical physics.
Although they originally appeared in the classical version
(as Hamiltonian structures on integrable hierarchies [\GD]),
the classical structure of these algebras has attracted a wider
attention only recently (see e.g.~[\cw]).
The classical structure of a \W\ is
expressed in terms of Poisson brackets between the generators.
For conventional Lie algebras, an alternative formulation in terms of
one-forms dual to the generators is available, which has proved useful,
especially in complex situations such as supergravity theories.
The equations expressing the algebraic structure are the \mc\ for the
left invariant one-forms on the group (see e.g.~[\CB]).
The \mc\ have a natural
generalization in which forms of arbitrary degree are admitted, and
which leads to \fda s
(for a review, see [\CDF]).
In this \letter\ I extend this concept
further, allowing the generators of the algebra as zero-forms in the \fda.
This generalization allows one to formulate nonlinear algebras in a
compact way.

An interesting aspect of this formulation is that the \mc\ are the
equations of motion of a three-dimensional action which generalizes
\cs\ theory for a compact gauge group. In the case of the \vir\
algebra, this action is related to Polyakov's chiral induced gravity
in exactly the same way as \cs\ theories are related to
WZW-models [\ik].

The relation with generalized \cs\ theory is very useful for the explicit
construction of \W s. As an example, I give the complete structure
of $N=2$ super $W_4$, for which the conventional construction has not been
carried out.

\begin{2.} The structure of an arbitrary Lie algebra
can be specified by the \mc\
$$dA+\half[A,A]=0\,,\eqno(2.1)$$
where  $[,]$ is the Lie bracket
and $A=A^aT_a$ is the canonical Lie algebra valued one-form.
The components $A^a$ of $A$ are left invariant one-forms on the
corresponding group, dual to the left invariant vector fields $T_a$,
which are in one to one correspondence with the generators of the algebra.
Equivalently, the left hand side of (2.1) can be viewed as the definition
of the curvature two-form corresponding to the gauge potential $A$.
The consistency condition
$d^2A=0$ is satisfied by virtue of the Jacobi identities satisfied by the
Lie bracket.

For the \vir\ algebra, the \mc\ can be written as
$$d\mu(x)+\mu(x)\mu'(x)=0\,,\eqno(2.2)$$
$$da+\intx\mu(x)\mu'''(x)=0\,.\eqno(2.3)$$
Here, $x$ is a coordinate on the circle, and $\mu(x)$ and $a$
are the components of $A$ dual (in the above sense) to the stress tensor
$T(x)$ and the central element, respectively.

One can construct an alternative algebraic structure
involving the dual objects $\mu$ and $T$,
which is equivalent to (2.2,3).
Consider the system
$$d\mu(x)+\mu(x)\mu'(x)=0\,,\eqno(2.4)$$
$$\nabla T(x)+\mu'''(x)=0\,.\eqno(2.5)$$
Here, $\nabla\phi\equiv d\phi+s\mu'\phi+\mu\phi'$ denotes the covariant
derivative of the tensor $\phi$ of spin $s$, the spin
of $\mu$ and $T$ being $-1$ and $2$ respectively.
It is easy to verify that (2.4,5) is integrable
in the sense that the consistency conditions $d^2T=d^2\mu=0$ are satisfied.
Hence, this system defines
a (generalized) \fda, with the generator $T$ appearing as a zero-form.
Moreover, since (2.5) is the \vir\ Ward identity, which by itself completely
characterizes the \vir\ algebra [\HV], (2.4,5) can be considered as
equivalent to the system (2.2,3).
In this case, the conventional description in terms of one-forms (2.2,3)
is available, but in the case of the \wthree\ the inclusion of generators
in the dual description cannot be avoided.

To pass from one of the systems (2.2,3), (2.4,5) to the other, it is useful
to notice that both are related to the \3al action
$$S=\intx\int (2TF+\mu\mu''')\,,\eqno(2.6)$$
where $F\equiv d\mu+\mu\mu'$ is the field strength corresponding to $\mu$.
The second integral here is over a two dimensional surface perpendicular
to the $x$ direction. The differential forms in this \letter\ have
components only along this surface.
The action (2.6) consists of a ''kinetic'' piece, which is the product of
the field strength $F$ and the generator $T$,
and the \gfc\ appearing in (2.3).
A procedure to derive this action from the system (2.2,3) is described
in detail elsewhere [\ik].
However, as I will discuss presently,
actions of this type can be constructed directly, without reference
to a system of the form (2.2,3).
Eqs.~(2.4,5), or generalizations thereof,
are then obtained as the \eom\ corresponding to this action.

The action (2.6) is the analogue for the Virasoro algebra
of the \cs\ action for a compact gauge group, which is well known to be
related to the corresponding \km\ algebra.
This analogy with \cs\ theory is clear if one writes the latter in
a $2+1$ decomposition
$$S=\hbox{tr}\intx\int (2A_xF-AA')\,.\eqno(2.7)$$
Under ordinary gauge transformations in three dimensions,
$A_x$ transforms as a current of the \km\ algebra, so that the first term in
(2.7) is analogous to the first term in (2.6). The second term in (2.7)
is the \km\ cocycle and is the analogue of the \gfc\ in (2.6).

The Virasoro action (2.6) can be generalized in an obvious manner to
include additional spins. For each additional generator, one also introduces
a dual one-form. The kinetic piece should then contain the
product of each generator with the corresponding field strength,
and the \gfc\
should be extended to the most general closed two-form
that can be constructed
from the dual one-forms and the generators. This procedure automatically
yields \eom\ which form an integrable system, and is in fact an efficient
method to construct such systems.

The \wthree\ algebra contains, in addition to the spin $2$ field $T(x)$
and the central element, a spin $3$ field $W(x)$. The corresponding dual
one-form $\nu(x)$ has spin $-2$. In order to incorporate these fields
in the system (2.2,3), one has to construct a closed two-form
involving $\nu$ analogous to the \gfc\ $\intx\mu\mu'''$.
The leading part of this two-form is expected to be
$$\intx\nu\n{5}\,.\eqno(2.8)$$
Such a cocycle however cannot be constructed from $\nu$ alone;
one also needs to include the stress tensor $T$.

To see this, I supplement eqs.~(2.4,5) with
$$\nabla\nu=0\,.\eqno(2.9)$$
The left hand side of this equation is the minimal
form of the field strength for $\nu$, as dictated by the
conformal symmetry. I will always take this minimal expression as
a starting point for constructing the action. The full field strength
then follows as an equation of motion of this action,
and will contain in general corrections to the ansatz of the form (2.9).

To construct the $W_3$ cocycle, I consider the exterior derivative of
$\intx\nu\n{5}$
$$d\intx\nu\n{5}=\intx(10\nu\nu'''\mu'''-6\nu\nu'\m{5})\,.\eqno(2.10)$$
Observe that the right hand side of (2.10) only involves derivatives
of $\mu'''$. Because of this, it can be cancelled by adding
the following terms
$$\intx(10\nu\nu'''T-6\nu\nu'T'')\,.\eqno(2.11)$$
The derivative of (2.11), in turn, is cancelled by a term
quadratic in $T$
$$\intx16\nu\nu'T^2\,,\eqno(2.12)$$
which is closed by itself. Hence the sum of eqs.~(2.8,11,12) is a closed
two-form, which can be added to the \gfc\ in eq.~(2.3).
It expresses the fact that the commutator of two $W$ charges
contains terms proportional to the stress tensor, although it does so in
an unconventional way by introducing terms proportional to $T$
in the central extension. The advantage of this approach is that the
nonlinearities can be incorporated naturally by allowing higher
powers of the generators, such as the $T^2$ term of eq.~(2.12).

The \vir\ \cs\ action (2.6) can now be extended to the following
$W_3$ \cs\ action
$$S=\intx\int (2TF+2W\nabla\nu+\mu\mu'''+\nu\n{5}+10\nu\nu'''T-6\nu\nu'T''
    +16\nu\nu'T^2)\,.\eqno(2.13)$$
The equations of motion corresponding to this action are
$$d\mu+\mu\mu'-\nu\nu'''-6\nu'\nu''+32\nu\nu'T=0\,,\eqno(2.14)$$
$$\nabla\nu=0\,,\eqno(2.15)$$
$$\nabla T+\mu'''+3\nu'W+2\nu W'=0\,,\eqno(2.16)$$
$$\nabla W+D^5\nu=0\,,\eqno(2.17)$$
where $D^5$ is the \gdo\ acting on spin $-2$
$$D^5\nu\equiv\n{5}+10\nu'''T+15\nu''T'+9\nu'T''+2\nu T'''
  +16\nu'T^2+16\nu TT'\,.\eqno(2.18)$$
This system is again easily seen to be integrable.
Notice that eqs.~(2.16,17) are the familiar $W_3$ Ward identities
(see e.g. [\SvN]).
These Ward identities, as well as the definitions of the curvatures
(2.14,15) are summarized in a compact way in the \cs\ action (12).

Other extensions of the \vir\ algebra can be constructed
in exactly the same way. For example, the $W_3^{(2)}$ algebra [\P],
containing, besides the stress tensor, two bosonic spins $3/2$, $G_\pm$,
and a $U(1)$ charge, $J$, is characterized by the following action
$$\eqalign{S=\intx\int &(2TF+G_+\nabla\psim-G_-\nabla\psip+2J\nabla A
 +\mu\mu'''+AA'+\psip\psim''\cr
 &\quad+2Q\psip\psim'J+Q\psip\psim J'+\half\psip\psim T
	+(Q^2-{\textstyle{1\over4}})\psip\psim J^2)\,.\cr}\eqno(2.19)$$
Here, the subscripts $\pm$ label the $U(1)$ charges of the fields
in units of $Q$, which is determined to be $\half\sqrt{-3}$,
and $\nabla\phi$ now denotes the covariant derivative including the $U(1)$
connection $A$ of the field $\phi$ of spin $s$ and charge $q$
$$\nabla\phi\equiv\phi+s\mu'\phi+\mu\phi'+qA\phi\,.\eqno(2.20)$$

\begin{3.} An interesting class of extensions of the \vir\ algebra are the
super \W s. The formalism presented in this \letter\ can be
extended to the super case, describing the algebra in terms of an
action for superfields in superspace. I will discuss here the
$N=2$ super $W_3$- [\LPRSW] and super $W_4$-algebras.

I will use the following notation. Let $\th\equiv\th_1+\th_2$,
$\bar\th\equiv\th_1-\th_2$ denote chiral combinations of the fermionic
coordinates, and $D\equiv\partial_{\bar\th}+\th\partial_x$,
$\D\equiv\partial_\th+\bar\th\partial_x$ the corresponding
superderivatives.
Then $\phi'=\half(\D D+D\D)\phi$ is the ordinary derivative,
while I define $\dot\phi\equiv\half(\D D-D\D)\phi$.

The super \vir\ algebra is characterized by the following action in $N=2$
superspace
$$S=\intxth(2JF+\mu\dot\mu')\,,\eqno(3.1)$$
where $\mu$ and $J$ now denote $N=2$ superfields having the corresponding
bosonic fields as their lowest components.
The $N=2$ field strength  is defined as
$$F\equiv d\mu+\mu\mu'+\half\D\mu D\mu\,,\eqno(3.2)$$
and, similarly, the $N=2$ supercovariant derivative is
$$\nabla\phi=d\phi+s\mu'\phi+\mu\phi'+\half\D\mu D\phi+\half D\mu\D\phi\,.
  \eqno(3.3)$$
Notice that the generator
dual to $\mu$ now is a spin $1$ charge $J$. This is dictated by the
dimension of the $N=2$ superspace measure. The super \vir\ Ward identity
following from (3.1) is
$$\nabla J+\dot\mu'=0\,.\eqno(3.4)$$
As an aside, the maximal
super extension of the \vir\ algebra that can be formulated
in this fashion is the $N=3$ algebra, with action
$$S=\intx d\th_1\,d\th_2\,d\th_3\,\int(2\Phi(d\mu+\mu\mu'+
     {\textstyle{1\over4}}D_i\mu D_i\mu)+\mu D_1D_2D_3\mu)\,,
  \eqno(3.5)$$
where $\Phi$ is a spin $1/2$ field.

A useful property of the $N=2$ superconformal algebra is the parity
symmetry $\th\leftrightarrow\bar\th$, $\mu\rightarrow\mu$, $J\rightarrow-J$.
Both the measure and the lagrangian in (3.1)
are odd under this parity symmetry.

\def\D{\bar D}
\def\5#1{#1^{(5)}}
\def\4#1{#1^{(4)}}
\def\half{{\textstyle{1\over2}}}
\def\n#1{\nu^{(#1)}}
\def\m#1{\mu^{(#1)}}
\def\D{\bar D}
\def\f#1/#2.{{\textstyle{#1\over#2}}}

\def\dJ{\dot J}
\def\dS{\dot S}
\def\dV{\dot V}
\def\ds{\dot \s}
\def\fv{\r\5{\s}}
\def\fo{\r\4{\s}}
\def\th{\r\s'''}
\def\tw{\r\s''}
\def\on{\r\s'}
\def\z{\r\s}
\def\dfi{\r\5{\ds}}

\def\dth{\r\ds'''}
\def\dtw{\r\ds''}
\def\don{\r\ds'}
\def\dz{\r\ds}

\def\Dfo{\r\4{\D\s}}
\def\Dth{\r\D\s'''}
\def\Dtw{\r\D\s''}
\def\Don{\r\D\s'}
\def\Dz{\r\D\s}
\def\pc{\hbox{p.c.}}
\def\rtw{\sqrt{-\f2/5.}}
\def\rth{\sqrt{\f3/5.}}
The $N=2$ super \wthree\ contains,
in addition to $\mu$ and $J$,
again a spin $-2$ one-form $\nu$ which is now accompanied by a
spin $2$ charge $S$. $\nu$ and $S$ have negative and positive parity,
respectively.
To determine the closed two-form to appear in the
action, I first determine the structure required by superconformal
covariance. That is, analogously to the bosonic case,
I take as a starting point the superconformal Ward identity (3.4)
supplemented by the equations
$$\nabla\nu=\nabla S+\dot\nu'''=0\,,\eqno(3.6)$$
and determine the most general cocycle that can be built from $\nu$
and $S$, consistent with parity.
In this case, there are two independent cocycles, one with leading term
$\nu\dot\nu'''$, and an additional one linear in $S$, with leading term
$\nu\dot\nu'S$. A short calculation analogous to the one leading to (2.11,12)
gives the full structure of these cocycles
\def\r{\nu}\def\s{\nu}
$$\eqalign{\omega_0=
&\half\dth+\th J+3\Dtw DJ+\don\dJ+4\Don DJ'\cr
&+\half\dz\dJ'+\f3/2.\Dz DJ''\cr
&-\half\don J^2-\Dz JDJ-\half\dz JJ'+2\on J\dJ-\f15/4.\on\D JDJ\cr
&-\half\Dz JDJ'+\f9/4.\Dz\dJ DJ-\f3/4. J'DJ\cr
&-\Dz J^2DJ-\on J^3-\pc\,,\cr
}\eqno(3.7)$$
\def\r{\nu}\def\s{\nu}
$$\eqalign{\omega_1=\al_1(
&\half\don S+\half\Don DS+\f1/4.\dz S'+\f3/20.\on\dS+\f3/10.\Dz DS'\cr
&+\f7/10.\on JS+\f9/10.\Dz SDJ-\f1/10.\Dz JDS)-\pc\cr
}\eqno(3.8)$$
\medskip\noindent Here p.c. denotes the parity conjugate.

The action generating the differential algebra can be written as
$$S=\intxth(2JF+2S\nabla\nu+\mu\dot\mu'+\om_0+\om_1)\,,\eqno(3.9)$$
involving a single constant $\al_1$. This constant is to be determined
by imposing the integrability condition on the equations of motion.
In this way one obtains $\al_1^2=-25$, where the sign ambiguity
corresponds to the freedom of flipping the sign of both $\nu$ and $S$.

The $N=2$ super \wfour\ contains, in addition to the fields
of the $N=2$ \wthree, a spin $-3$ one-form $\pi$ of positive parity,
and a spin 3 charge $V$ of negative parity.
As a starting point, one supplements eqs.~(3.4,6) with
$$\nabla\pi=\nabla V+\dot\pi^{(5)}=0\,.\eqno(3.10)$$
In addition to (3.7,8), one has then the following closed
combinations
\def\s{\pi}\def\r{\pi}
$$\eqalignno{\omega_2=\hskip.7mm
&\half\dfi+\f3/2.\fv J+\f15/2.\Dfo DJ+5\dth\dJ+\f15/2.\fo J'\cr
&+20\Dth DJ'+\f15/2.\dtw\dJ'+15\th J''+\f45/2.\Dtw DJ''+\f9/2.\don\dJ''\cr
&+12\Don DJ'''+\don\dJ'''-\f15/2.\on\4{J}+\f5/2.\Dz D\4{J}\cr
&-\f5/2.\dth J^2+5\Dth JDJ-\f15/2.\dtw JJ'-\f105/4.\th\D JDJ\cr
&+15\th J\dJ+\f195/4.\Dtw \dJ DJ-\f15/2.\Dtw JDJ'-\f45/4.\Dtw J'DJ\cr
&-\f3/2.\don\D J'DJ+8\don\dJ^2-\f9/2.\don J'^2-\f9/2.\don JJ''\cr
&-\f9/2.\Don JDJ''+\f99/2.\Don\dJ'DJ-9\Don J''DJ+64\Don\dJ DJ'\cr
&-12\Don J'DJ'\cr
&-\f3/4.\dz\D J''DJ+8\dz\dJ\dJ'-\dz JJ'''-3\dz J'J''\cr
&-9\on J\dJ''-\f45/2.\on J'\dJ'+\f105/4.\on\D J''DJ+35\on\D J'DJ'\cr
&-\Dz JDJ'''-\f15/4.\Dz J'DJ''-5\Dz J''DJ'-\f5/2.\Dz J'''DJ\cr
&+15\Dz\dJ''DJ+\f65/2.\Dz\dJ'DJ'+\f95/4.\Dz\dJ DJ''&(3.11)\cr
&-\f15/2.\th J^3-\f45/2.\Dtw J^2DJ-8\don J^2\dJ+\f3/4.\don J\D JDJ\cr
&-\f61/2.\Don J^2DJ'-\f177/4.\Don JJ'DJ-\f67/4.\Don J\dJ DJ\cr
&-4\dz J^2\dJ'-8\dz J\dJ J'+\f3/8.\dz J'\D JDJ+\f3/4.\dz J\D JDJ'\cr
&+24\on J\dJ^2+\f63/2.\on JJ'^2+9\on J^2J''\cr
&-\f735/8.\on\dJ\D JDJ-\f17/4.\on J\D JDJ'\cr
&+\f225/8.\Dz\dJ^2DJ-\f105/8.\Dz J'^2DJ-\f17/2.\Dz J\dJ'DJ-13\Dz JJ''DJ\cr
&-\f25/2.\Dz\dJ J'DJ-\f33/4.\Dz J\dJ DJ'-\f121/4.\Dz JJ'DJ'\cr
&-\f23/2.\Dz J^2DJ''-\f5/2.\Dz DJ'\D JDJ\cr
&+2\don J^4+4\Don J^3DJ\cr
&+4\dz J^3J'-24\on J^3\dJ+46\on J^2\D JDJ\cr
&+4\Dz J^3DJ'-26\Dz J^2\dJ DJ+14\Dz J^2J'DJ\cr
&+6\on J^5+6\Dz J^4DJ-\pc\,,\cr
&\cr
\global\def\r{\nu}\global\def\s{\pi}
\omega_3=\al_3(
&\half\th S+\f3/4.\Dtw DS+\f3/2.\tw S'+\f3/20.\don \dS+\f6/5.\Don DS'\cr
&+\f3/2.\on S''+\f1/10.\dz \dS'+\f1/2.\Dz DS''+\half\z S'''\cr
&-\f3/10.\don JS-\f9/10.\Don SDJ-\f3/20.\Don JDS\cr
&-\f1/5.\dz J'S-\f1/5.\dz JS'-\f1/8.\dz \D JDS\cr
&+\on\dJ S+\f9/20.\on J\dS+\f21/8.\on\D JDS\cr
&-\f1/10.\Dz JDS'-\f1/8.\Dz J'DS+\f5/8.\Dz\dJ DS&(3.12)\cr
&-\half\Dz SDJ'-\f3/4.\Dz S'DJ+\half\Dz\dS DJ\cr
&+\f3/10.\z J\dS'+\z\dJ S'+\f9/20.\z J'\dS\cr
&+\half\z\dJ'S+\f7/4.\z\D J'DS+\f21/10.\z\D JDS'\cr
&-\f7/5.\on J^2S-\f3/10.\Dz J^2DS-\f4/5.\Dz JDJS-\f11/10.\z J^2S'\cr
&-2\z JJ'S+\f1/5.\z J\D JDS\cr
&+\f3/40.\nu\nu'V+\f1/40.\nu\D\nu DV)-\pc\,,\cr
&\cr
\global\def\r{\pi}\global\def\s{\pi}
\omega_4=\al_4(
&\half\dth S+\half\Dth DS+\f3/4.\dtw S'+\f3/20.\th\dS+\f9/10.\Dtw DS'\cr
&\f9/20.\don S''+\f3/5.\Don DS''+\f1/10.\dz S'''-\f3/35.\on\dS''+
 \f1/7.\Dz DS'''\cr
&+\f6/5.\th JS+\f21/5.\Dtw DJS-\f3/10.\Dtw JDS\cr
&+\f8/5.\don\dJ S-\f3/10.\don\D JDS\cr
&+\f29/5.\Don SDJ'-\f3/10.\Don JDS'+\f3/20.\Don\dS DJ\cr
&+\f7/4.\Don\dJ DS+\f81/20.\Don S'DJ-\f9/20.\Don J'DS\cr
&+\f4/5.\dz\dJ'S+\f4/5.\dz\dJ S'-\f3/20.\dz\D J'DS-\f3/20.\dz\D JDS'\cr
&+\f6/35.\on J''S-\f267/140.\on J'S'-\f51/70.\on
JS''+\f39/70.\on\dJ\dS&(3.13)\cr
&-\f13/14.\on\D J'DS-\f12/35.\on\D JDS'\cr
&+\f31/14.\Dz SDJ''-\f3/35.\Dz JDS''+\f11/140.\Dz\dS DJ'
 +\f29/28.\Dz\dJ DS'\cr
&+\f79/28.\Dz S'DJ'-\f33/140.\Dz J'DS'+\f3/35.\Dz\dS'DJ+\f25/28.\Dz\dJ'DS\cr
&+\f33/28.\Dz S''DJ-\f5/28.\Dz J''DS\cr
&-\f4/5.\don J^2S-\f4/5.\Don J^2DS-\f8/5.\Don JDJS-\f2/5.\dz J^2S'
 -\f4/5.\dz JJ'S\cr
&-\f39/140.\on J^2\dS+\f258/70.\on J\dJ S-\f969/140.\on\D JDJS
 -\f17/28.\on J\D JDS\cr
&-\f59/70.\Dz JDJ'S-\f17/35.\Dz J^2DS'+\f135/28.\Dz\dJ DJS
 -\f163/140.\Dz J'DJS\cr
&-\f5/14.\Dz J\dJ DS-\f11/14.\Dz JJ'DS-\f1/28.\Dz JDJ\dS
 -\f11/14.\Dz JDJS'\cr
&+\f25/56.\Dz\D JDJDS\cr
&-\f129/70.\don J^3S-\f153/70.\Dz J^2DJS+\f6/35.\Dz J^3DS\cr
\global\def\r{\nu}
&+\f3/10.\don V+\f2/5.\Don DV+\f1/5.\dz V'+\f1/7.\on\dV\cr
&+\f2/7.\Dz DV'+\f3/28.\z\dV'\cr
&+\f33/70.\on JV+\f9/14.\Dz DJV -\f2/35.\Dz JDV\cr
&+\f81/140.\z J'V+\f39/140.\z JV'-\f9/20.\z\D JDV)-\pc\,,\cr
&\cr
\global\def\r{\pi}\global\def\s{\pi}
\omega_5=\al_5(
&\half\th V+\half\Dtw DV+\f1/14.\don\dV\cr
&+\f4/7.\Don DV'+\f1/28.\dz\dV'-\f2/7.\on V''+\f5/28.\Dz DV''\cr
&-\f3/14.\don JV-\f3/14.\Don VDJ-\f3/14.\Don JDV\cr
&-\f3/28.\dz J'V-\f3/28.\dz JV'+\f13/7.\on\dJ V+\f3/14.\on J\dV
 -\f57/28.\on\D JDV\cr
&-\f1/28.\Dz VDJ'-\f3/28.\Dz JDV'+\f5/8.\Dz\dJ DV-\f9/56.\Dz J'DV\cr
&+\f15/56.\Dz\dV DJ-\f9/56.\Dz V'DJ&(3.14)\cr
&-\f11/7.\on J^2V-\f5/7.\Dz JDJV-\f2/7.\Dz J^2DV)-\pc\,,\cr
&\cr
\global\def\r{\pi}\global\def\s{\pi}
\omega_6=\al_6(
&\half\don S^2+\Don SDS+\half\dz SS'+\half\on S\dS+\Dz SDS'\cr
&+\half\on JS^2+\half\Dz DJS^2)\cr
+\be_6(
&6\Don SDS+5\dz SS'+5\on S\dS+\half\on\D SDS\cr
&+12\Dz SDS'+3\Dz S'DS&(3.15)\cr
&-4\on JS^2-6\Dz DJS^2+2\Dz JSDS)\cr
+\ga_6(
&-4\Dz SDS'+5\Dz S'DS+\Dz\dS DS\cr
&+8\Dz DJS^2-4\Dz JSDS)-\pc\,,\cr
&\cr
\global\def\r{\pi}\global\def\s{\pi}
\omega_7=\al_7(
&\half\on SV+\f1/4.\Dz DSV)\cr
+\be_7(
&\half\on SV+\f1/6.\Dz SDV)-\pc\,,&(3.16)\cr
&\cr
\global\def\r{\nu}\global\def\s{\pi}
\omega_8=\al_8(
&\half\on S^2+\f1/4.\Dz SDS+\f3/4.\z SS')-\pc&(3.17)\cr
}$$
\medskip\noindent The action is the sum of the standard
kinetic piece and the sum of these cocycles
$$S=\intxth(2JF+2S\nabla\nu+2V\nabla\pi+\mu\dot\mu'+
  \sum\om_i)\,.\eqno(3.18)$$
Again, the constants are determined by imposing the
integrability conditions on the equations of motion.
One obtains
$$\halign to \hsize
{\qquad#\hfil&\qquad#\hfil&\qquad#\hfil\tabskip=0pt plus
10cm&\hfil#\tabskip=0pt\cr
	   $\al_1=7\rtw$,&&&\cr
	   $\al_3=-12\rth$,&&&\cr
	   $\al_4=28\rtw$,&&&\cr
	   $\al_5=\f14/5.\rth$,&&&(3.19)\cr
	   $\al_6=-46$,&$\be_6=10$,&$\ga_6=5$,&\cr
	   $\al_7=-100\rtw\rth$,&$\be_7=60\rtw\rth$,&&\cr
	   $\al_8=-60\rtw\rth$.&&&\cr
	   }$$
\smallskip\noindent The sign ambiguity of $\rtw$ ($\rth$) corresponds to
the freedom of flipping the sign of $\nu$ and $S$ ($\pi$ and $V$).

\begin{4.} In this \letter\ I have presented an extension of the concept of
free differential algebras which allows one to incorporate nonlinearities
of the type encountered in \W s. The \mc\
are the equations of motion corresponding to a \cs\ type action
which encodes the algebraic structure in a compact way. This formalism
is very well suited to the explicit construction of more complicated
algebras, and I have used it here
to determine the complete structure
of the $N=2$ super $W_4$-algebra.

After writing this \letter\ I became aware of [\SVN],
in which curvatures for nonlinear algebras are defined
containing coadjoint scalars. This generalization of curvatures
is equivalent to
the generalization of free differential algebras considered here.

\begin{Acknowledgements:} I would like to thank K.~Pilch for useful
comments on the ma\-nus\-cript, and L.~Romans and N.~Warner
for their interest in this work.
This research was partially supported by
the Department of Energy Contract \#DE-FG03-84ER40168.

\def\eef #1 {\llap{#1\enspace}\ignorespaces}
\def\ref{\beginsection{References}\frenchspacing
    \parindent=0pt\leftskip=1truecm\parskip=8pt plus 3pt
    \everypar={\eef}}
    \def\npb#1{Nucl. Phys. {\bf B#1}}
    \def\cmp#1{Commun. Math. Phys. {\bf#1}}

    \def\plb#1{Phys. Lett. {\bf #1B}}
    \def\ijmp#1{Int. J. Mod. Phys. {\bf#1}}

    \def\mpl#1{Mod. Phys. Lett. {\bf A#1}}
    \def\faa#1{Funct. Anal. Appl. {\bf#1}}
    
    \def\ibid#1{ibid. {\bf#1}}

\def\hb{\hfil\break}

{\ref

[1] A.B. Zamolodchikov, Teo. Mat. Fiz. {\bf 65} (1985) 347;
\hb V.A. Fateev and S.L. Luk'yanov, \ijmp{A3} (1988) 507

[2] I.M. Gel'fand and L.A. Dickey, Russ. Math. Surv. {\bf 30} (1975) 77;
    \faa{10} (1976) 4; \ibid{11} (1977) 93

[3] A. Bilal and J.-L. Gervais, \plb{206} (1988) 412; \npb{314} (1989) 646;
\hb I. Bakas, \plb{213} (1988) 313; \npb{302} (1988) 189;
\hb P. Di Francesco, C. Itzykson and J.-B. Zuber, \cmp{140} (1991) 543;
\hb J.M. Figueroa-O'Farril and E. Ramos, \cmp{145} (1992) 43;
    \npb{368} (1992) 361; \plb{282} (1992) 357;
\hb V.A. Fateev and S.L. Luk'yanov, \ijmp{A7} (1992) 853;
\hb J.-L. Gervais, {\it Classical $A(N)$ W Geometry}, LPTENS-91-35

[4] Y. Choquet-Bruhat, C. DeWitt-Morette with M. Dillard-Bleick,
    {\it Analysis, Manifolds and Physics}, North-Holland 1982

[5] L. Castellani, R. D'Auria and P. Fr\' e, in
    {\it Supergravity and Supersymmetry '83}, ed. B. Milewski,
    World Scientific 1983

[6] R.E.C. Perret, in preparation

[7] H. Verlinde, \npb{337} (1990) 652

[8] H. Ooguri, K. Schoutens, A. Sevrin and P. van Nieuwenhuizen,
    \cmp{146} (1992) 616

[9] A.M. Polyakov, \ijmp{A5} (1990) 833;
\hb M. Bershadsky, \cmp{139} (1991) 71

[10] H. Lu, C.N. Pope, L.J. Romans, X. Shen and X.-J. Wang, \plb{264}
    (1991) 91;
\hb L.J. Romans, \npb{369} (1992) 403;
\hb K. Huitu and D. Nemeschansky, \mpl{6} (1991) 3179

[11] K. Schoutens, A. Sevrin and P. van Nieuwenhuizen, \npb{349} (1991) 791;
    \plb{255} (1991) 549

}

\vfil\eject\end